\documentstyle[sprocl,amsmath,epsfig]{article}
\newcommand{\ksl}{k \! \! \! /}
\newcommand{\Omegasl}{\Omega \! \! \! \! /}
\newcommand{\Deltasl}{\Delta \! \! \! \! /}

\begin{document}
\pagestyle{empty}

\begin{flushleft}
\large
{SAGA-HE-124-97, SLAC-PUB-7626
\hfill Sept. 2, 1997}  \\
%%%\hfill \today}  \\
\end{flushleft}
 
\vspace{1.5cm}
 
\begin{center}
 
\Large{{\bf One and two loop anomalous dimensions }} \\
\vspace{0.2cm}

\Large{{\bf for the chiral-odd structure function $\bf h_1$}} \\

\vspace{1.0cm}
 
\Large
{S. Kumano $^{a),\, b)}$ and M. Miyama $^{a)}$ }         \\
 
\vspace{0.4cm}
  
\end{center}

\Large
\noindent
{a) Department of Physics, Saga University}      \\
 
\vspace{-0.6cm}
\noindent
\ \ \ \ 
{Honjo-1, Saga 840, Japan $^*$} \\

\vspace{-0.3cm}
\noindent
{b) Stanford Linear Accelerator Center} \\
 
\vspace{-0.6cm}
\noindent
\ \ \ \ 
{Stanford University, Stanford, CA 94309, U.S.A. $^\dagger$} \\

\begin{center}
\vspace{0.3cm}
 
\large
{Invited talk at the 10th Summer School \& Symposium on Nuclear} \\

\vspace{0.2cm}
{ Physics, ``QCD, Lightcone Physics and Hadron Phenomenology"}

\vspace{0.2cm}
{Seoul National University, Seoul, Korea} \\

\vspace{0.2cm}
{June 23 - June 28, 1997 (talk on June 26, 1997)} \\

\end{center}
 
\vspace{0.3cm}

\vfill
 
\noindent
{\rule{6.cm}{0.1mm}} \\
 
\vspace{-0.5cm}
\normalsize
\noindent
{* Email: kumanos@cc.saga-u.ac.jp, 96td25@edu.cc.saga-u.ac.jp.} \\
\vspace{-0.2cm}
\noindent
{\ Information on their research is available at http://www.cc.saga-u.ac.jp}  \\

\vspace{-0.24cm}
\noindent
{\ \ /saga-u/riko/physics/quantum1/structure.html.} \\

\vspace{-0.0cm}
\noindent
{$\dagger$ Work partially supported by the US Department
           of Energy under the contract} \\ 

\vspace{-0.4cm}
\noindent
{\ \ DE--AC03--76SF00515.} \\

\vspace{+0.0cm}
\hfill
{to be published in proceedings by the World Scientific}

\vfill\eject
\setcounter{page}{1}
\pagestyle{plain}
%%%%%%%%%%%%%%%%%%%%%%%%%%%%%%%%%%%%%%%%%%%%%%%%%%%%%%%%%%%%%%%%%%%%%%
%%%%%%%%%%%%%%%%%%%%%%%%%%%%%%%%%%%%%%%%%%%%%%%%%%%%%%%%%%%%%%%%%%%%%%

\title{One and two loop anomalous dimensions \\
       for the chiral-odd structure function $\bf h_1$}

\vspace{-0.8cm}

\author{S. Kumano $^{a), b)}$ and M. Miyama $^{a)}$}

\address{a) Department of Physics, Saga University \\
         Honjo-1, Saga 840, Japan \\
         b) Stanford Linear Accelerator Center \\
         Stanford University, Stanford, CA 94309, U.S.A.}

\vspace{-0.6cm}
\maketitle\abstracts{
Because the chiral-odd structure function $h_1$ will be measured
in the polarized Drell-Yan process, it is important to predict the behavior
of $h_1$ before the measurement. In order to study the $Q^2$ evolution of $h_1$,
we discuss one and two loop anomalous dimensions which are calculated
in the Feynman gauge and minimal subtraction scheme. 
}
  
\vspace{-0.8cm}
%%%%%%%%%%%%%%%%%%%%%%%%%%%%%%%%%%%%%%%%%%%%%%%%%%%%%%%%%%%%%%%%%%%%%%%%%%%%%%%%
\section{Introduction}

\vspace{-0.3cm}
Despite much effort to understand the proton spin structure,
we have not reached a consensus in interpreting it
in terms of quark and gluon spins.
As another way to investigate the proton spin,
the transversity distribution $h_1$ is proposed.
It will be measured in transversely polarized Drell-Yan processes
at RHIC. Before the experimental data are taken, we had better
predict the behavior of $h_1$ as much as we can. Some model estimates
on the $x$ dependence have been already done, for example, in
the MIT bag model. Furthermore, the leading-order (LO) splitting function
and anomalous dimensions were already calculated.\cite{lo} 
Therefore, rough behavior is already known although 
they subject to experimental tests.

In these days, the next-to-leading (NLO) analyses are the standard
in studying parton distributions not only in the unpolarized ones
but also in the longitudinally polarized ones. 
Hence, it is very important to understand the NLO splitting function
or equivalently anomalous dimensions for $h_1$.
Here, we discuss the one and two loop results, 
which had been completed recently.\cite{nlo}
The bare operator is defined by
$O_B^n =\overline\psi_B Q_B^n \psi_B =Z_{O^n} O_R^n$ with 
the renormalized one $O_R^n$.
Once the renormalization constant $Z_{O^n}$ is determined,
the anomalous dimension is calculated by
$\gamma_{_{O^n}} = \mu \, \partial (\ln Z_{O^n})/\partial\mu$.
We discuss the calculations of $\gamma_{_{O^n}}$
for chiral-odd operators
in the Feynman gauge and minimal subtraction scheme.

\vspace{-0.2cm}
%%%%%%%%%%%%%%%%%%%%%%%%%%%%%%%%%%%%%%%%%%%%%%%%%%%%%%%%%%%%%%%%%%%%%%%%%%%%%%%%
\section{One-loop anomalous dimensions for $\bf h_1$}

\begin{figure}[h]
\parbox[b]{0.46\textwidth}{
   \begin{center}
      \epsfig{file=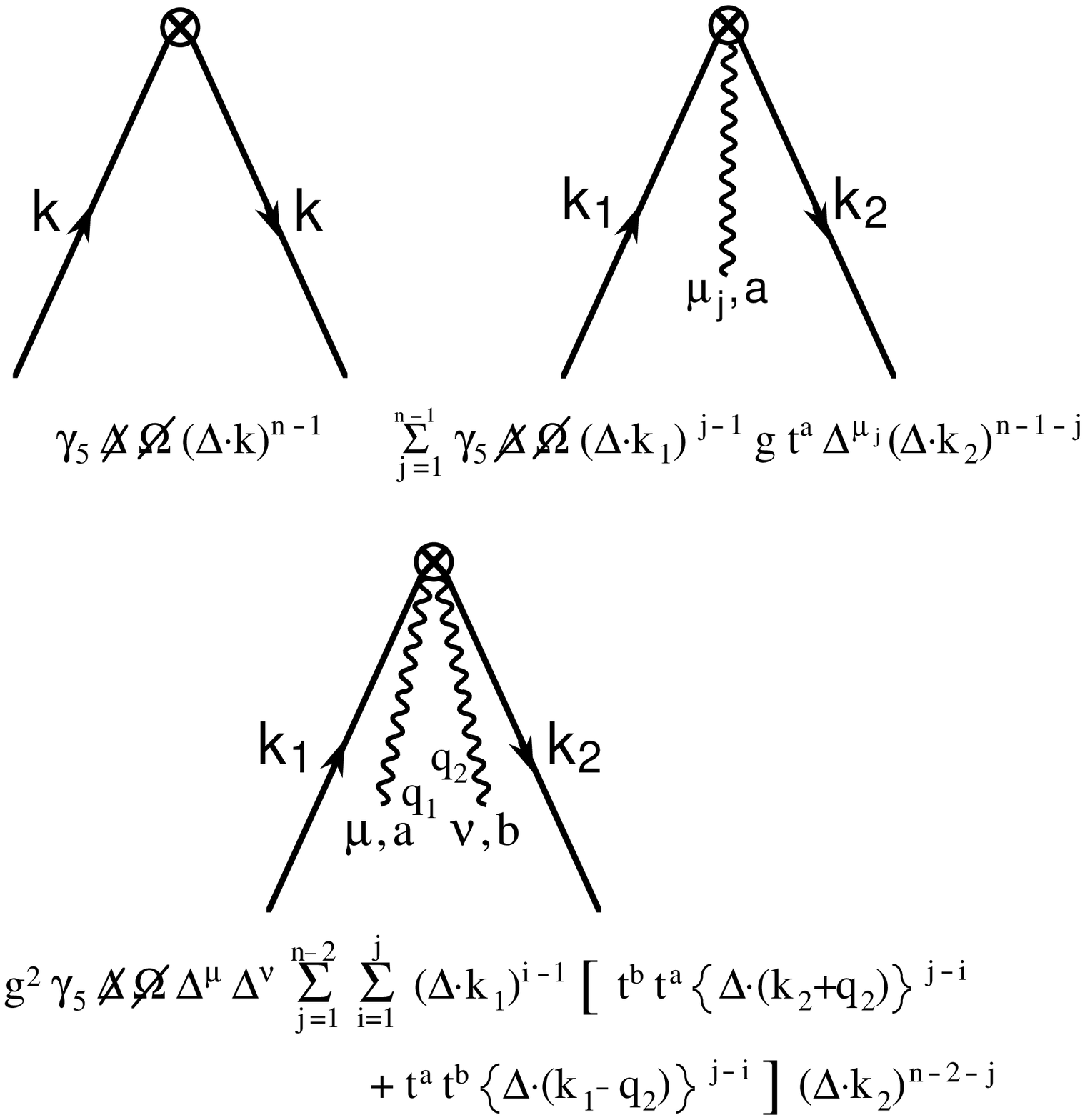,width=5.0cm}
   \end{center}
 \vspace{-0.2cm}
   \caption{\footnotesize Feynman rules.}
   \label{fig:feyn}
}\hfill
\parbox[b]{0.46\textwidth}{
   \begin{center}
      \epsfig{file=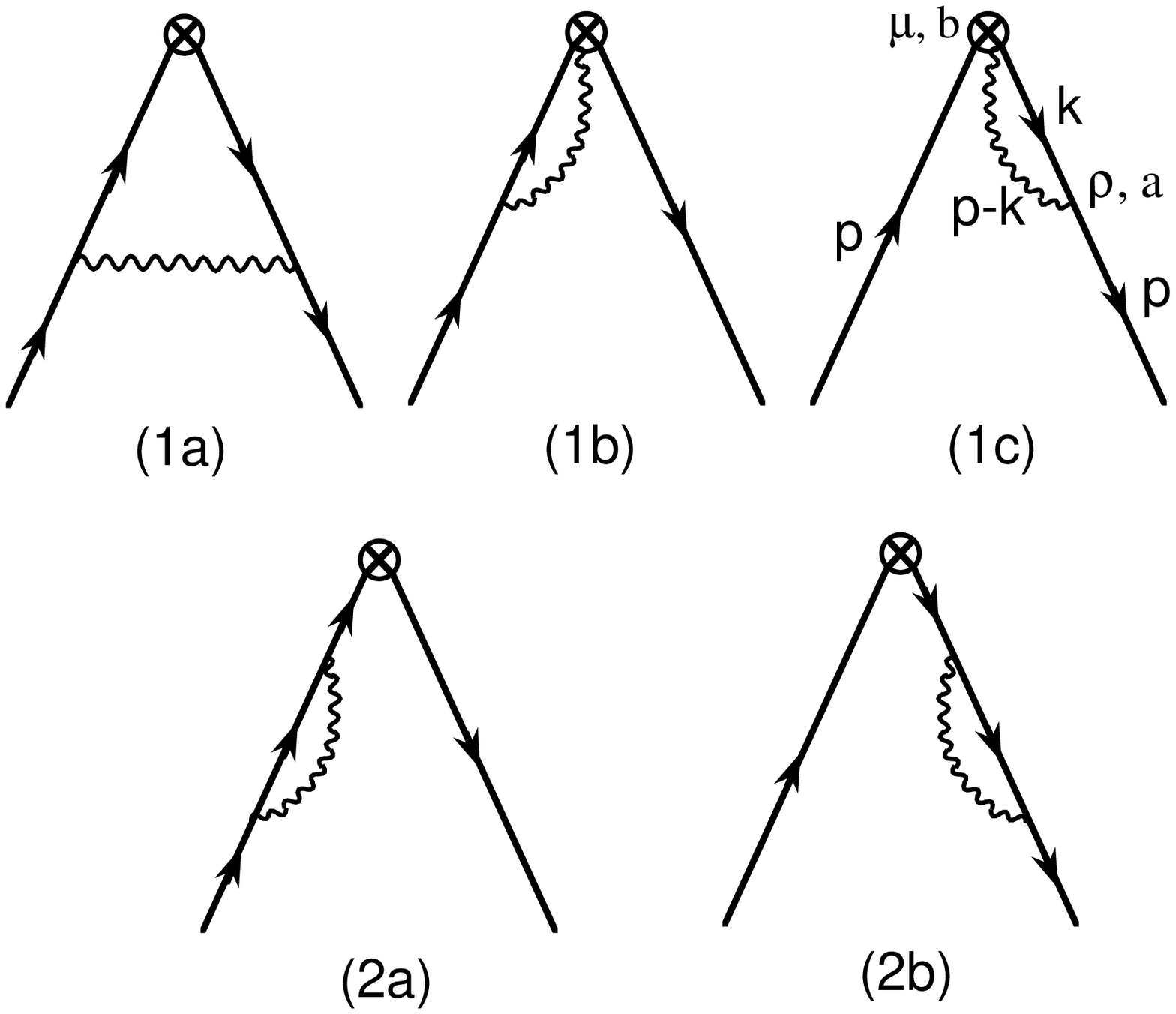,width=5.5cm}
   \end{center}
 \vspace{-0.2cm}
\caption{\footnotesize One-loop diagrams.}
\label{fig:1loop}
}
\end{figure}

\vspace{-0.3cm}
Because the gluon field does not contribute to the anomalous
dimensions directly due to the chiral-odd nature of $h_1$,
the calculation becomes simpler.
In order to study $h_1$ in perturbative QCD, 
we need to introduce a set of local operators ($n=1, 2, \cdot\cdot\cdot$)
\begin{equation}
O^{\nu\mu_1 \cdot\cdot\cdot \mu_n}
   = S_n \, \overline\psi \, i \, \gamma_5 \, \sigma^{\nu\mu_1} \, 
         i D^{\mu_2} \cdot\cdot\cdot i D^{\mu_n} \, \psi
     - \ {\rm trace} \ {\rm terms}
\ .
\label{ope}
\end{equation}
Here, $S_n$ symmetrizes the Lorentz indices $\mu_1$, ..., $\mu_n$,
and $iD^\mu=i\partial^\mu + g t^a A_a^\mu$ is the covariant derivative.
For calculating the anomalous dimensions of these operators,
Feynman rules at the operator vertices should be provided.
In order to satisfy the symmetrization condition and to remove
the trace terms, the tensor 
$\Delta_{\mu_1} \Delta_{\mu_2} \cdot\cdot\cdot \Delta_{\mu_n}$
with the constraint $\Delta^2=0$ is multiplied.
However, the operators in Eq. (\ref{ope}) are associated with one-more
Lorentz index $\nu$, so that it is convenient to introduce
another vector $\Omega_\nu$ with the constraint $\Omega\cdot\Delta$=0.
Then, Feynman rules with zero, one, and two gluon vertices become
those in Fig. \ref{fig:feyn}.\cite{nlo}

\vspace{-0.6cm}
%%%%%%%%%%%%%%%%%%%%%%%%%%%%%%%% figure %%%%%%%%%%%%%%%%%%%%%%%%%%%%%%%%%%%%%%
\noindent
%%%%%%%%%%%%%%%%%%%%%%%%%%%%%%%% figure %%%%%%%%%%%%%%%%%%%%%%%%%%%%%%%%%%%%%%
\vspace{-0.4cm}

The one-loop diagrams are shown in Fig. \ref{fig:1loop}.
The anomalous dimension $\gamma_n$ can be calculated
from the $1/\epsilon$ singularity in the dimensional regularization
with the dimension $d=4-\epsilon$.
Evaluating the diagram (1a), we find that there is no such singularity.
It means that there is no contribution to $\gamma_n$ from (1a).
Obviously, contributions from the diagrams (1b) and (1c) are the same.
We show the calculation of the (1c) diagram. With the Feynman rules
in Fig. \ref{fig:feyn}, the diagram is evaluated as
\begin{align}
I_{(1c)} & = \int \frac{d^dk}{(2\pi)^d}
       \, ig t^a \gamma^\rho \, \frac{i \ksl}{k^2}
       \, \frac{-ig_{\rho\mu}\delta_{ab}}{(p-k)^2}
       \, \sum_{j=1}^{n-1} \, \gamma_5 
       \, \Deltasl \, \Omegasl \, (\Delta\cdot p)^{j-1}
       \, g t^b \Delta^{\mu} \, (\Delta\cdot k)^{n-1-j}
\nonumber \\
         & = \gamma_5 \Deltasl \, \Omegasl \, (\Delta\cdot p)^{n-1}
       \, \biggl [ -\frac{1}{\epsilon} \frac{4g^2}{(4\pi)^2}
       C_F \, \left ( \frac{-p^2}{4\pi} \right )^{-\epsilon/2}
       \sum_{j=1}^{n-1} \frac{1}{j+1} \biggl ]
\ ,
\end{align}
where $C_F$ is given by $C_F=(N_c^2-1)/(2N_c)$ with the number of color $N_c$.
From this equation, the diagram (1c) contribution to the
anomalous dimension becomes
$\gamma_n^{(1c)} = \gamma_n^{(1b)} =  4 \, C_F
       \, \sum_{j=1}^{n-1} \, 1/(j+1)$.
Throughout this paper, we show the anomalous dimension multiplied by
the factor $(4\pi)^2/g^2$ in the one-loop case and
$(4\pi)^4/g^4$ in the two-loop one.
Adding all the one-loop contributions in Fig. \ref{fig:1loop},
we obtain \cite{lo}
\begin{eqnarray}
\gamma_n^{(0)} = 2 \, C_F \biggl [ 
       \, 1 + 4 \, \sum_{j=1}^{n-1} \frac{1}{j+1} \ \biggl ]
\ .
\end{eqnarray}
The corresponding LO splitting function is given in Appendix.

\vspace{-0.4cm}
%%%%%%%%%%%%%%%%%%%%%%%%%%%%%%%%%%%%%%%%%%%%%%%%%%%%%%%%%%%%%%%%%%%%%%%%%%%%%%%%
\section{Two-loop anomalous dimensions}
\vspace{-0.3cm}

\begin{figure}[h]
\parbox[b]{0.46\textwidth}{
   \begin{center}
      \epsfig{file=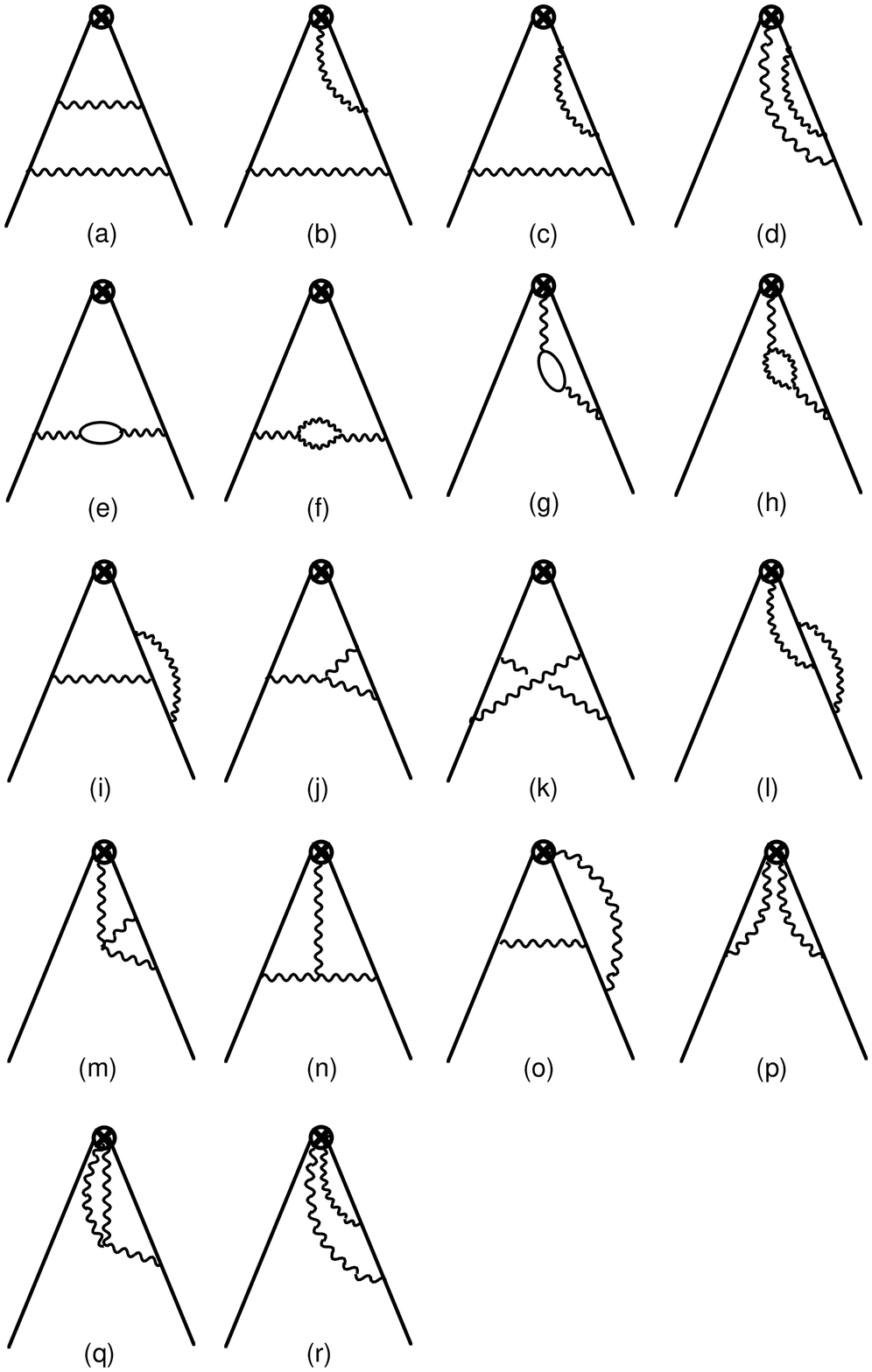,width=4.0cm}
   \end{center}
 \vspace{-0.2cm}
   \caption{\footnotesize Two-loop contributions.}
\label{fig:2loop}
}\hfill
\parbox[b]{0.46\textwidth}{
   \begin{center}
      \epsfig{file=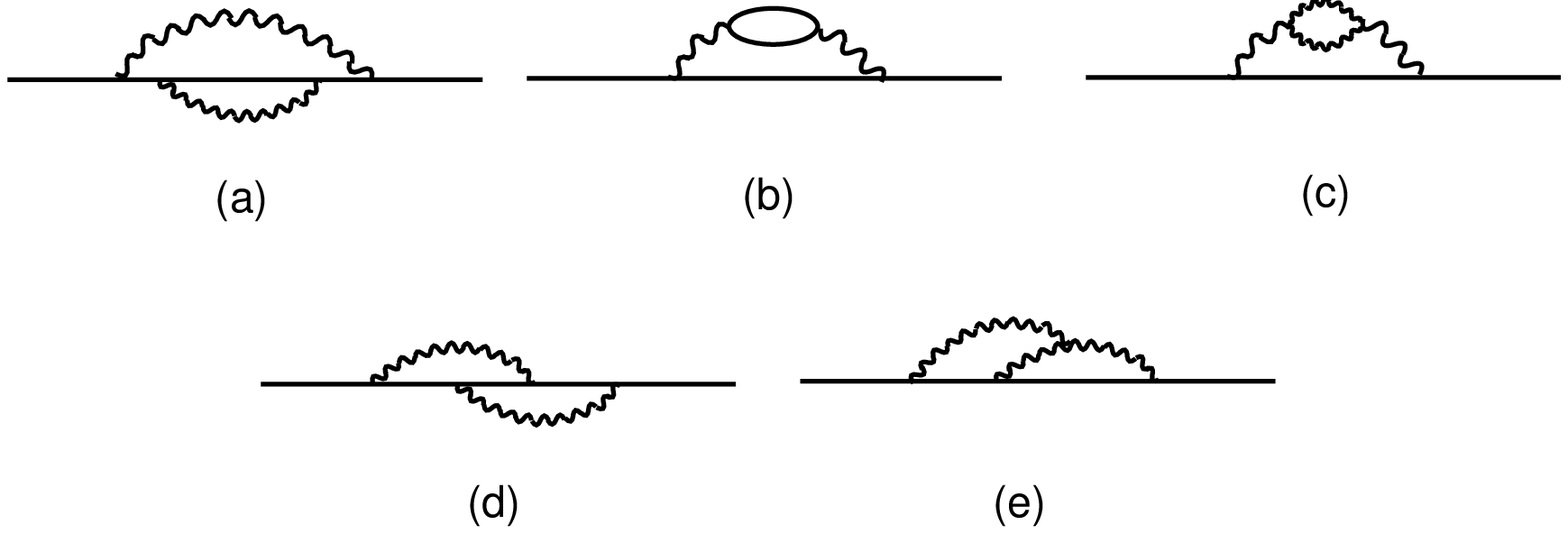,width=4.7cm}
   \end{center}
 \vspace{-0.2cm}
\caption{\footnotesize Quark field renormalization.}
\label{fig:quark}
\vspace{0.3cm}
   \begin{center}
      \epsfig{file=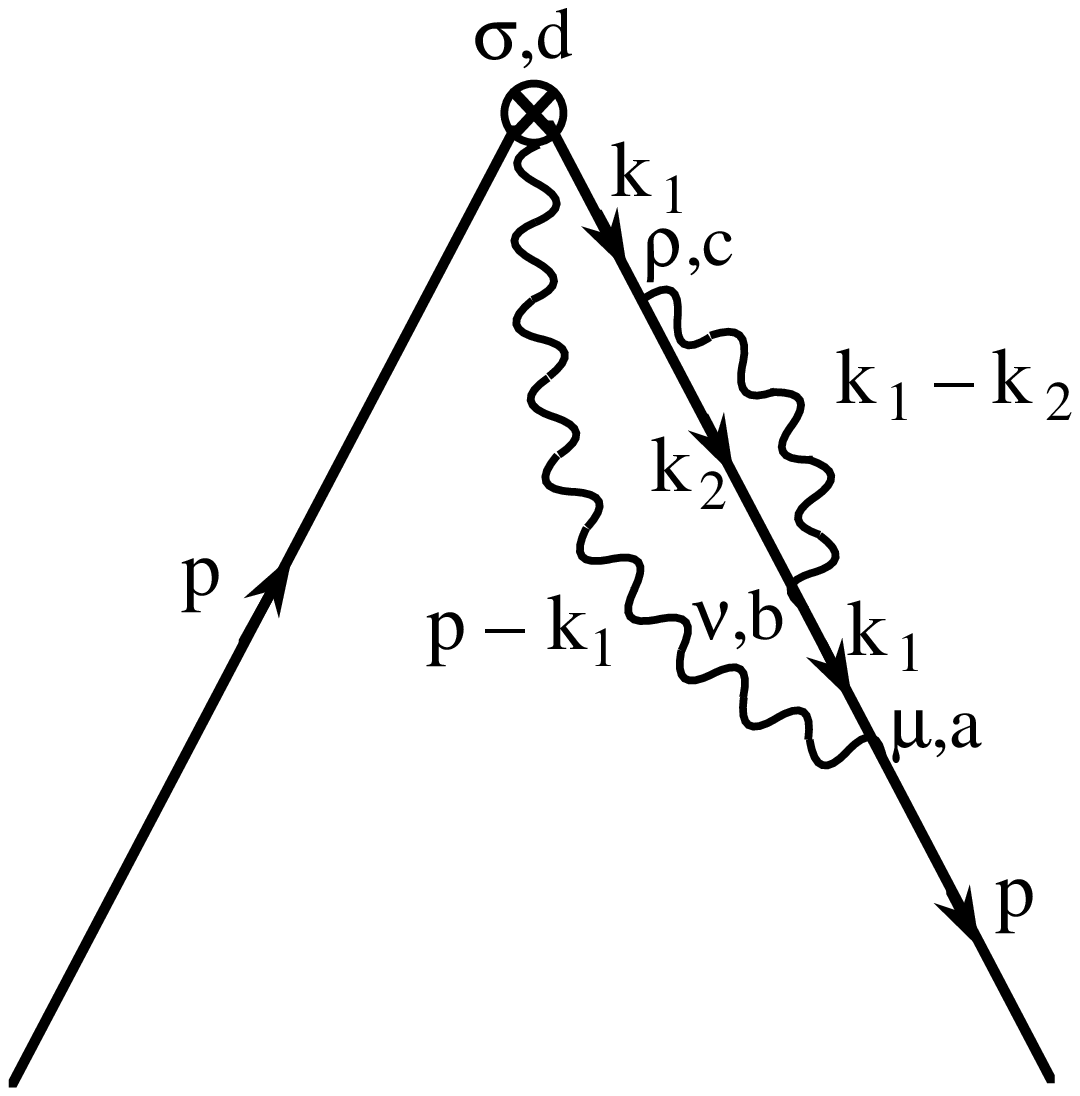,width=2.8cm}
   \end{center}
 \vspace{-0.2cm}
\caption{\footnotesize Lorentz and color indices.}
\label{fig:d}
}
\end{figure}

The two-loop anomalous dimensions were calculated recently.\,\cite{nlo}
Because the gluon field does not contribute directly, the calculation
is similar to the nonsinglet one.\cite{frs}
All the two-loop contributions are shown in Figs. \ref{fig:2loop} and
\ref{fig:quark}.

\vspace{-0.7cm}
%%%%%%%%%%%%%%%%%%%%%%%%%%%%%%%% figure %%%%%%%%%%%%%%%%%%%%%%%%%%%%%%%%%%%%%%
\noindent
%%%%%%%%%%%%%%%%%%%%%%%%%%%%%%%% figure %%%%%%%%%%%%%%%%%%%%%%%%%%%%%%%%%%%%%%
\vspace{-0.2cm}

\noindent
The quark-field renormalization in Fig. \ref{fig:quark} was already 
calculated, so that the problem is to calculate the contributions
in Fig. \ref{fig:2loop}. 
Because it is too complex to explain all the calculations, we discuss
only a simple example.
The Lorentz and color indices of the diagram (d) in Fig. \ref{fig:2loop}
are shown in Fig. \ref{fig:d}, and it is given by
\begin{align}
I_{(d)} = & \int \frac{d^dk_1}{(2\pi)^d} \frac{d^dk_2}{(2\pi)^d}
       \, ig t^a \gamma^\mu \, \frac{i \ksl_1}{k_1^2} 
       \, ig t^b \gamma^\nu \, \frac{i\ksl_2}{k_2^2} 
       \, \frac{-ig_{\nu\rho}\delta_{bc}}{(k_1-k_2)^2} 
       \, ig t^c \gamma^\rho \, \frac{i \ksl_1}{k_1^2} 
       \, \frac{-ig_{\mu\sigma}\delta_{ad}}{(p-k_1)^2}
\nonumber \\
    &  \times \, \, \sum_{j=1}^{n-1} \, \gamma_5 \Deltasl \, \Omegasl
       \, \, (\Delta\cdot p)^{j-1} 
       \, g t^d \Delta^\sigma \, (\Delta\cdot k_1)^{n-1-j}
\ .
\label{eqn:d}
\end{align}
Subtracting a one-loop counter term from the above equation
and evaluating singular terms, we have
\begin{equation}
I_{(d)}'= \gamma_5 \Deltasl \, \Omegasl \, \, (\Delta\cdot p)^{n-1}
      \frac{2g^4}{(4\pi)^4}  C_F^2  
      \sum_{j=1}^{n-1} 
      \frac{1}{j+1} \left [ - \frac{2}{\epsilon^2}
         + \frac{1}{\epsilon} \, \{ 1-S_1 (j+1) \} \right ]
\ ,
\end{equation}
where $S_1(n)=\sum_{k=1}^n 1/k$. From the above equation, 
$\gamma_n$ becomes 
$\gamma_n^{(d)} =  8 \, C_F^2 \, [ \, G_1(n) - S_1(n) \, ] $
with $G_1(n)=\sum_{j=1}^n (1/j) \sum_{i=1}^j (1/i)$.
The factor of two is included by considering a similar diagram
with gluons attached to the initial quark line.
This result is exactly the same as the unpolarized one in Ref. 3.
The reason is the following.
Because the operator vertex part $\gamma_5 \Deltasl \, \Omegasl \, $
can be separated from the integrals in Eq. (\ref{eqn:d}), 
the renormalization constant
is independent of the operator form.
Therefore, if the gluon lines are attached 
only to the final-quark or initial-quark line, the anomalous 
dimensions are the same as those of the unpolarized.
In this way, we do not have to repeat the same calculations 
for the diagrams (d), (g), (h), (l), (m), (q), and (r).
The results are listed in Ref. 3.
Furthermore, calculating the integrals, we find easily that
there is no contribution from the diagrams (a), (b), (c), and (p).
The problem reduces to calculations of remaining diagrams (e), (f), 
(i), (j), (k), (n), and (o). It is tedious to calculate some of
these diagrams. Because the calculation is rather lengthy, we do
not show each calculation. Adding all the contributions in
Figs. \ref{fig:2loop} and \ref{fig:quark}, we obtain
the final result for the two-loop anomalous dimensions as \cite{nlo}
\begin{align}
\gamma_n^{(1)} =& C_F^2 \, \bigg [ \, 32 \, S_1(n) \, 
             \{ \, S_2(n) \, - \, S_2'(n/2) \, \} 
            \, + \, 24 \, S_2(n) 
            \, - \, 8 \, S_3'(n/2) \, + \, 64 \, \widetilde S(n) 
\nonumber \\
&
\ \ \ \ 
            \, - \,  8 \, \frac{1-(-1)^n}{n(n+1)} \, - \, 3 \, \, \bigg ]
          + \, C_F \, T_R \, \bigg [ \, -\, \frac{160}{9} \, S_1(n)
         \, + \, \frac{32}{3} \, S_2(n) \, + \, \frac{4}{3} \, \, \bigg ]
\nonumber \\
&         + \, C_F \, C_G \, \bigg [ \, \frac{536}{9} \, S_1(n)
             \, - \, \frac{88}{3} \, S_2(n) \, + \, 4 \, S_3'(n/2)
             \, - \, 32 \, \hat S(n) 
\nonumber \\
& 
\ \ \ \
           - \, 16 \, S_1(n) \, \{ \,2 S_2(n) \, - \, S_2'(n/2) \, \}
         \, + \, 4 \, \frac{1-(-1)^n}{n(n+1)} \, - \, \frac{17}{3}
        \, \, \bigg ]
\ .
\end{align}
The corresponding NLO splitting function is given in Appendix.

\vspace{-0.2cm}
%%%%%%%%%%%%%%%%%%%%%%%%%%%%%%%%%%%%%%%%%%%%%%%%%%%%%%%%%%%%%%%%%%%%%%%%%%%%%%%%
\section{Conclusion}
\vspace{-0.3cm}

The two-loop anomalous dimensions had been calculated recently,
so that the next-to-leading-order analyses became
possible not only for unpolarized and $g_1$ structure functions
but also for the structure function $h_1$.

\vspace{-0.2cm}
%%%%%%%%%%%%%%%%%%%%%%%%%%%%%%%%%%%%%%%%%%%%%%%%%%%%%%%%%%%%%%%%%%%%%%%%%%%%%%%%
\section*{Acknowledgment}
\vspace{-0.24cm}

S.K. would like to thank the organizers of this summer school for
financial support for his participation.

\vspace{-0.2cm}
%%%%%%%%%%%%%%%%%%%%%%%%%%%%%%%%%%%%%%%%%%%%%%%%%%%%%%%%%%%%%%%%%%%%%%%%%%%%%%%%
\section*{Appendix: Splitting functions}
\vspace{-0.15cm}

The LO and NLO splitting functions are \cite{nlo}
\begin{align} 
\Delta_T P_{qq}^{(0)} (x) &= C_F \left[ \frac{2x}{(1-x)_+}+\frac{3}{2}
\delta (1-x) \right] \ ,
\label{eqn:plo}
\\
\Delta_T P_{qq,\pm}^{(1)} (x) &\equiv \Delta_T P_{qq}^{(1)} (x) \pm 
\Delta_T P_{q\bar{q}}^{(1)} (x) \ ,
\label{eqn:pnlo1}
\\
\Delta_T P_{qq}^{(1)} (x)     &= C_F^2 \Bigg[ 1-x - \left( \frac{3}{2}
            + 2 \ln (1-x) \right) \ln x \; \; \delta_T P_{qq}^{(0)}(x) 
\nonumber \\
& \! \! \! \! \! \! \! \! \! \! \! \! \! \! \! \! \! \! 
  \! \! \! \! \! \! \! \! \!
                  + \left. \left( \frac{3}{8} -\frac{\pi^2}{2} + 6\zeta (3) 
                           \right) \delta (1-x) \right] 
                 + \frac{1}{2} C_F N_C \bigg [ - (1-x) 
\nonumber \\
& \! \! \! \! \! \! \! \! \! \! \! \! \! \! \! \! \! \! 
  \! \! \! \! \! \! \! \! \!
  \left.
                               + \left( \frac{67}{9} + \frac{11}{3} \ln x
                               + \ln^2 x - \frac{\pi^2}{3} \right) 
                                        \delta_T P_{qq}^{(0)}(x)  
                  + \left( \frac{17}{12} + \frac{11 \pi^2}{9} 
                  - 6 \zeta (3) \right) \delta (1-x) \right] 
\nonumber \\
& \! \! \! \! \! \! \! \! \! \! \! \! \! \! \! \! \! \! 
  \! \! \! \! \! \! \! \! \!
                               +\frac{2}{3} C_F T_f  
                    \left[ \left( - \ln x -\frac{5}{3} \right) \delta_T 
           P_{qq}^{(0)}(x) - \left( \frac{1}{4} + \frac{\pi^2}{3} \right) 
           \delta (1-x) \right] \ ,
\label{eqn:pnlo2}
\\
\Delta_T P_{q\bar{q}}^{(1)} (x) &= C_F \left( C_F - \frac{1}{2} N_C \right)
\Bigg[ -(1 -x) + 2 S_2 (x) \, \delta_T P_{qq}^{(0)}(-x) \Bigg] \ , 
\label{eqn:pnlo3}
\end{align}
where $\delta_T P_{qq}^{(0)} (x) = 2x/(1-x)_+$ and
$S_2(x) = \int_{x/(1+x)}^{1/(1+x)} dz/z \ln [(1-z)/z]$.

\vspace{-0.2cm}
%%%%%%%%%%%%%%%%%%%%%%%%%%%%%%%%%%%%%%%%%%%%%%%%%%%%%%%%%%%%%%%%%%%%%%%%%%%%%%%%

%%%%%%%%%%%%%%%%%%%%%%%%%%%%%%%%%%%%%%%%%%%%%%%%%%%%%%%%%%%%%%%%%%%%%%%%%%%%%%%%

\end{document}